# Roman CCS White Paper

# Asteroseismic sounding of bulge globular clusters with the Roman Space Telescope

**Roman Core Community Survey:** *Galactic Bulge Time Domain Survey*
**Scientific Categories:** *stellar physics and stellar types; stellar populations and the interstellar medium*
**Additional scientific keywords:** *Globular star clusters; Variable stars; Stellar evolution*

**Submitting Author:**

**Name: László Molnár**
Affiliation: Konkoly Observatory, ELKH CSFK, Budapest
Email: molnar.laszlo@csfk.org

**List of contributing authors** (including affiliation and email):

Name: Csilla Kalup
Affiliation: Konkoly Observatory, ELKH CSFK, Budapest
Email: kalup.csilla@csfk.org

Name: Meridith Joyce
Affiliation: Konkoly Observatory, ELKH CSFK, Budapest
Email: meridith.joyce@csfk.org

**Abstract:** *Globular clusters are relics of the early Universe and they hold clues to many aspects of stellar and galactic evolution. We propose to point the Roman Space Telescope at one or more clusters either as a part of or as an extension of the Galactic Bulge Time Domain Survey. This would provide a unique opportunity to apply the powerful toolkit of asteroseismology to a globular cluster, an observation that is largely out of reach for any other time-domain photometric missions. In this white paper we present the possible targets in the vicinity of the notional survey fields. Potential science cases include precise determination of stellar parameters throughout the cluster, accurate estimation of the integrated mass loss for metal-poor and metal-rich clusters, asteroseismic analysis and mass estimation for RR Lyrae stars, and determination of the seismic ages of clusters. We provide comparisons with other photometric missions and recommendations for maximizing the scientific return from a dedicated globular cluster observing run.*

1. Introduction

Globular clusters are dense agglomerates of old stars with thousands to millions of members in a bound system. Since they underwent only a limited amount of chemical evolution shortly after their formation, they are the relics of the early Universe, and can tell us about stellar evolution over 10+ billion years (see, e.g., Gatton et al. 2019; Krumholz et al. 2019).

Most of the Milky Way globular clusters can be found in the inner halo and the vicinity of the bulge. This can make their observations difficult, if they lay towards dense stellar fields or are obscured by dust in the disk of the Milky Way. They are also quite far away from the Sun, with even the closest one, Messier 4, being over 2000 pc away (Baumgardt & Vasiliev, 2021). These attributes, coupled with being intrinsically very dense stellar fields to observe, made it difficult to observe their variable star populations in great detail. Targeted studies from the ground undoubtedly revealed a great deal of information on various classes of variables, such as RR Lyrae stars, but the true seismic revolution is just beginning for globular clusters.

The first detailed seismic analysis of red giants in globular clusters came from the K2 mission of the Kepler space telescope. Kepler observed eight clusters, including M4, the only one for which results have been published. The cluster was searched for new variables thoroughly, resulting in the detection of millimagnitude RR Lyrae stars, among others (Wallace et al. 2019a,b). Solar-like oscillations detected in red giant, asymptotic giant and red horizontal branch stars allowed us to constrain stellar masses through asteroseismology and to estimate the amount of mass loss between evolutionary stages (Howell et al. 2022; Tailo et al. 2022). While analysis of further clusters like M80 and NGC5897 are underway, all of them are fainter than M4 and therefore offer much more limited possibilities for asteroseismology.

A temporal coverage akin to the K2 mission, coupled with a depth far exceeding that, therefore would create an unprecedented opportunity for the Nancy Grace Roman Space Telescope to thoroughly sound more globular clusters with asteroseismic methods. In this white paper we explore the various science cases, potential targets, and suggested modifications or extensions to the Galactic Bulge Time Domain Survey that can maximize the scientific return from pointing the telescope towards globular clusters.

## 2. Candidate targets

The notional field-of-view of the Galactic Bulge Time Domain Survey does not include any globular clusters. However, there are multiple potential targets in the vicinity of those, within 1-5 degrees (Fig. 1). Baade's Window includes two large, bright clusters, **NGC6522 and NGC6528** that are within 20' to each other, and therefore would easily fit into a single Roman pointing (Fig. 2). These are the highest-priority targets we propose. On the other side of Baade's Window, we have **NGC6540**, another bright cluster. We also identified four smaller clusters in the vicinity of Baade's Window: **Djorgovski 1 and 2**, as well as **Terzan 6 and 9**.

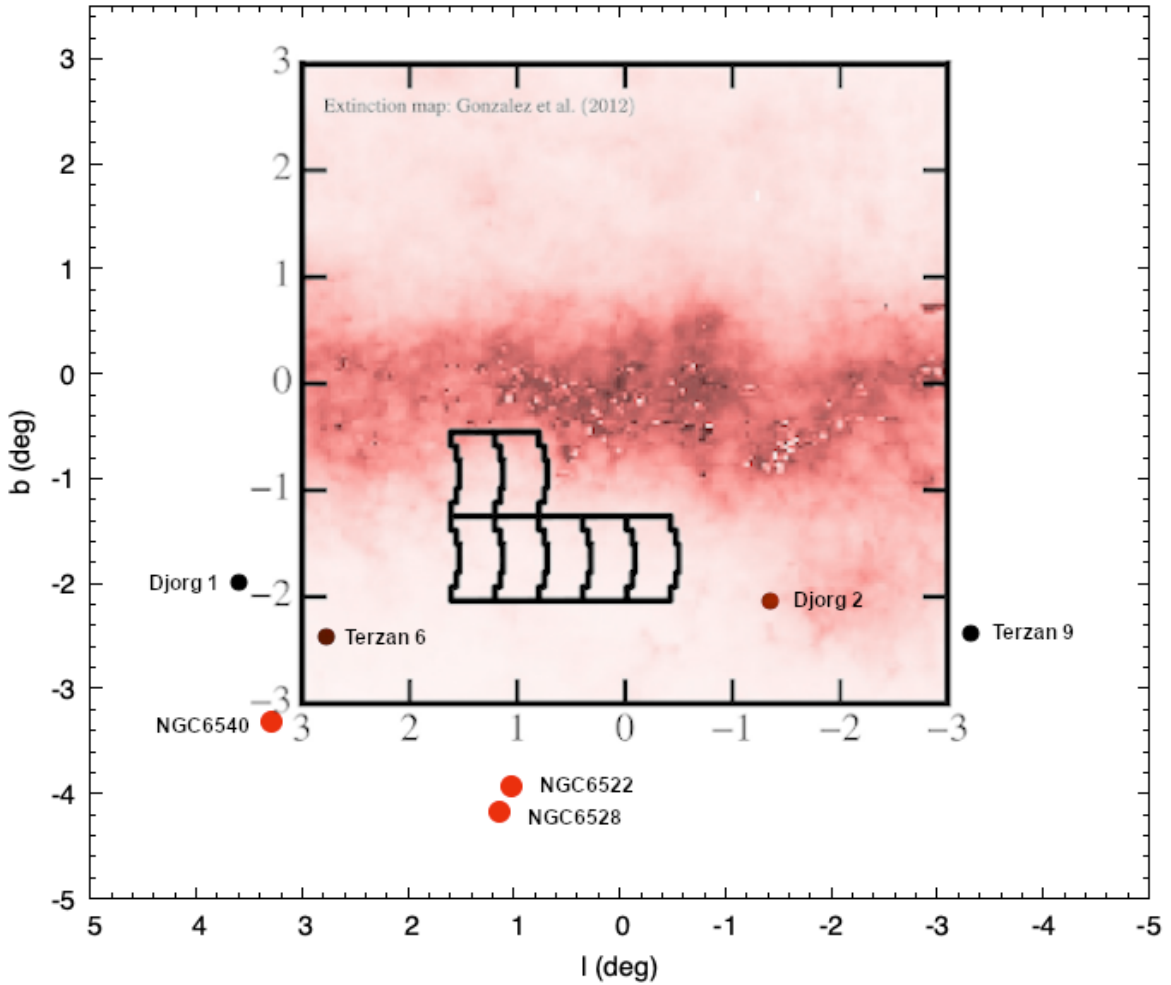

Figure 1: *positions of the proposed targets relative to the notional fields of the Survey. Original image: Penny et al. (2019).*

A further benefit of observing NGC6522 and NGC6528 simultaneously is that Roman would be able to sample two clusters with very different metallicities. NGC6522 is a moderately metal-poor object with [Fe/H] = -1.0 (Barbuy et al., 2021). In contrast, NGC6528 is one of the most metal-rich clusters around the Milky Way, with [Fe/H] =

-0.14 (Muñoz et al, 2018). This is also evident from the CMD of the cluster which features a very short and red Horizontal Branch, almost reminiscent of a Red Clump (Lagioia et al., 2015). The Horizontal Branch in NGC 6528 occurs at K=13.2 mag (Davidge, 2000).

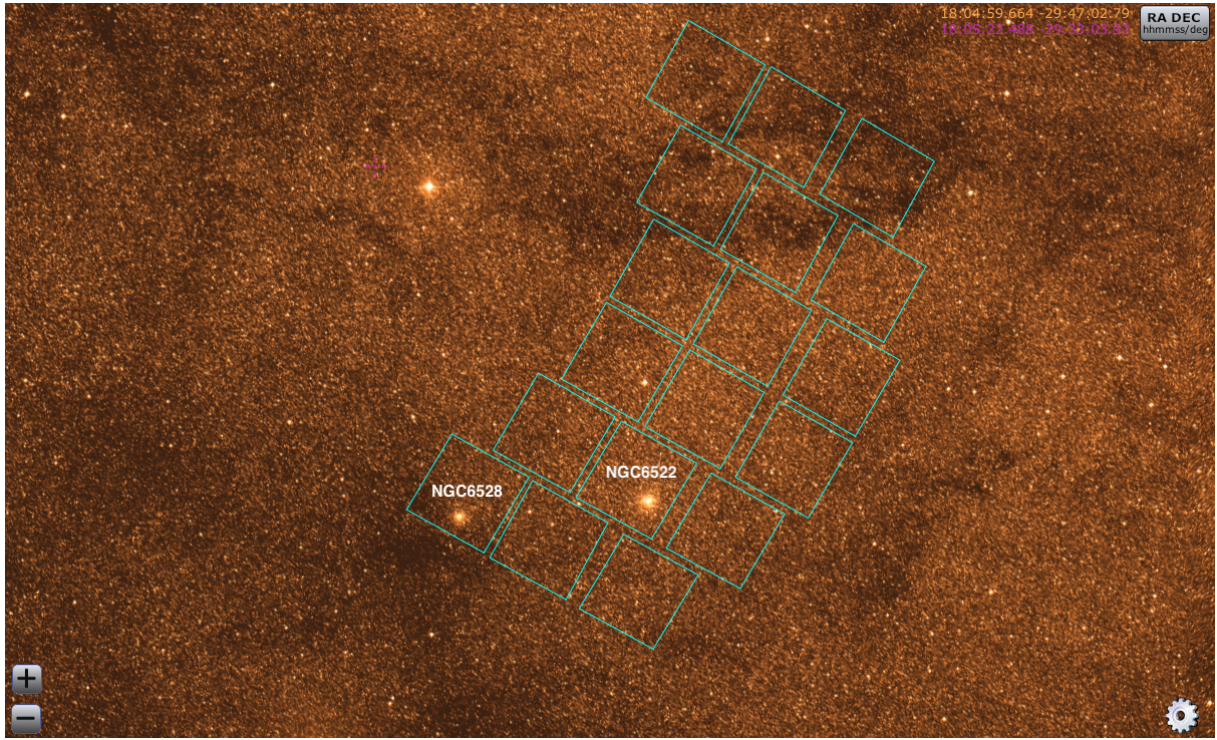

Figure 1. *The field-of-view of the Wide Field Imager, placed in an orientation that would cover both bright clusters in one pointing. Generated with the Field of View Overlay for Roman web tool provided by MAST.*

The average RR Lyrae and Main Sequence turnoff point brightness levels for NGC6522 are 16.7 and 19.7 mag in the F625W passband (Kerber et al., 2018), and slightly higher for NGC65288. This means that the RR Lyrae stars will be at the saturation limit for this mission.

NGC6540 first was classified as an open cluster, however, further observations showed a very compact cluster with a metallicity of [Fe/H] ≈ -1.0, which indicated that it is a globular cluster, with an extended horizontal branch and a poorly populated red giant branch. These characteristics suggest that it is a post-core-collapse cluster (Bica et al., 1994). It has a peculiar structure with a dense elongated nucleus in the north-west direction, while two clumps of bright stars extend from the cluster in the east-west direction. The only available color-magnitude diagram of NGC6540 was published by Bica et al. (1994), which shows a clear blue Horizontal Branch at V ≈ 15.9 mag.

Djorgovski 1 is one of the first globular clusters that was identified as a halo intruder, projected in the central part of the bulge, based on Hubble and Gaia data. Its color-magnitude diagram obtained using HST was deep and accurate enough to reveal the blue horizontal branch around 20 mag in the F606W passband. The main sequence turnoff point brightness is approximately at 24 mag, also in F606W (Ortolani et al., 2019a).

Djorgovski 2 is one of the closest globular clusters to the Galactic center, and has a moderate metallicity of [Fe/H] ≈ -1.0. It has a blue horizontal branch and is similar to the very old inner bulge globular clusters, therefore seems to be part of the primeval formation stages of the Milky Way. This cluster was observed by VLT and also by HST, which provide accurate color-magnitude diagrams, showing that the Horizontal Branch lies at 16.5 in the *I* band and at 18 mag in F606W, respectively (Ortolani et al., 2019b).

Terzan 5 and 9 are faint clusters embedded into interstellar dust and gas clouds in the vicinity of the galactic center. There is only one published color-magnitude diagram from Terzan 5, in infrared bands (Liu et al, 1994). Because of the proximity of the bulge, these clusters are heavily obscured by interstellar dust, therefore optical observations are difficult or even impossible.

These clusters offer different science goals. Bright clusters will allow Roman to survey their red giant populations for asteroseismic signals all the way down to the base of the RGB in high definition, allowing for detailed studies. For fainter clusters, general characterisation as well as ensemble analysis of the red giant, asymptotic giant and horizontal branches of the CMD are the starting points.

3. Science goals

Here we summarize some important science cases Roman could help answer if globular clusters are included in the Galactic Bulge Time Domain Survey.

**A full asteroseismic profile**

Roman will have the unique potential to survey globular clusters in depth in more ways than one. Not only will it provide an extremely deep but static image of the clusters through high-resolution imagery and multicolor brightness measurements but it will be able to collect highly detailed time-domain information as well. It will give us a lifetime opportunity to survey these old clusters through asteroseismology. Ensemble asteroseismology of open clusters observed by Kepler provided a plethora of new findings (see, e.g., Corsaro et al., 2012; Miglio et al. 2012), but even the oldest open clusters are considerably younger than the globular clusters around the Milky Way.

Asteroseismology has become a key tool in stellar physics. Since oscillations probe the stellar interior, they can provide parameters like age, mass and stellar radius more precisely than other types of measurements (Aerts 2021). Mode frequencies provide metrics that are much less sensitive to the stellar surface and thus to assumptions made for atmospheric boundary conditions. This, however, requires reliable detection of individual mode frequencies.

Roman will be able to survey the red giant branches of the clusters with the nominal survey cadence. Since the RGBs of most clusters will be close or even beyond the saturation limit of Roman, we expect very precise brightness measurements. This in turn suggests that red giant data will be sufficient for peak-bagging (i.e., detection of individual mode frequencies), including the detection of mixed modes. If Roman will detect mixed modes in globular cluster red giants, that will inform us about the properties of the cores of these very old stars (Mosser et al. 2014).

Roman could possibly have the photometric sensitivity to reach even lower and potentially detect solar-like oscillations on the subgiant branch for the brighter clusters. However, that will require a faster cadence than the proposed 15 min sampling, since solar-like oscillations scale strongly with the radius and thus are fastest on the main sequence.

On the other end of the color-magnitude diagram, Roman can have the power to disentangle the seismic signatures of AGB and RGB stars. Although at different evolutionary stages, AGB and RGB stars at the same luminosity look remarkably similar. This is also true for their seismic profiles, where differences can be detected through acoustic glitches that probe the deep interior (Dréau et al. 2021, Lindsay et al. 2023). In a globular cluster, however, we can disentangle the two branches, and we can compare their ensemble seismic signatures that differentiate stars through asteroseismology. However, this will require the handling of very bright stars (relative to Roman's sensitivity). We will discuss this technical challenge later.

**Globular cluster ages and seismic isochrones**

Ages of globular clusters are normally determined via isochrone fitting. However, these age values are sensitive to more than the quality and intrinsic spread of the photometric data the isochrones are applied to. Modeling choices, such as surface boundary conditions, convective parameters (especially the mixing length; see Joyce & Tayar, 2023), elemental abundances can inject large uncertainties into the isochrones themselves. This, in turn, has a large effect on the accuracy to which we are able to determine ages for old stars, with realistic uncertainties being in the 2-5 Gyr range (Joyce et al. 2022).

If Roman will have the sensitivity and cadence to detect seismic signatures, especially near the turnoff point in a bright cluster, we will have a chance to lower these uncertainties. Seismic ages will not be independent from classical ages, because they all come from stellar evolutionary models. Nevertheless, using mode frequencies instead of surface data (brightness and color) will lower the degrees of freedom considerably, which in turn will lower the uncertainties. For this, a grid of seismic isochrones will have to be computed: isochrones that include not only luminosities in various passbands but synthetic frequency spectra for every time step as well. And Roman could potentially have the power and sensitivity to produce observational data to which these seismic isochrones could be calibrated.

**Integrated mass loss**

Mass loss is a large uncertainty factor in stellar evolutionary models. Although we have empirical relations to estimate mass loss over the life of a star (see., e.g, Reimers 1975; Schröder & Cuntz, 2005) the details are still debated. When does it really happen? Is it gradual or episodic? What are the driving mechanisms?

Clusters provide a unique way to observationally measure the integrated mass loss, i.e., the total mass lost between evolutionary stages (Miglio et al. 2012). Since globular clusters are very old, the times required to go through the late evolutionary stages (RGB-HB and HB-AGB transitions) are very short in comparison to the ages of the stars. This translates to negligible initial mass differences between the stars populating the various branches. Thus, asteroseismic masses can be used directly to estimate the amount of mass loss between evolutionary stages (Howell et al. 2022; Tailo et al. 2022). With the inclusion of the metal rich cluster NGC6528, we can also test how integrated mass loss depends on metallicity.

Roman will be able to provide new constraints on stellar mass loss in different stellar populations through asteroseismic detections in multiple clusters. It will also offer a precision that far exceeds the capabilities of the Kepler mission, and which will have few if any counterparts in the foreseeable future (see Section 4).

**Classical pulsators**

Seismic masses can still be inferred from the HB using a new technique as well. Space-based photometry revealed a plethora of low-amplitude extra modes in RR Lyrae stars. While the origins of most of these modes are still uncertain, one set of modes detected in overtone (RRc or RR1) stars have a plausible explanation. The model presented by Dziembowski (2016) attributes these signals to high-degree, $\ell = 8$ and $9$ modes. The mode content of these stars can be modeled with pulsation codes (Netzel et al. 2022, and in prep). Thus we can extend the field of asteroseismology to include

RR Lyrae stars, which can serve as a mass estimate for the horizontal branch mass of the cluster, separate from red HB stars.

Mass estimates for RR Lyrae stars are valuable in their own right as well. Since very few binary RR Lyrae stars are known, accurate dynamical masses have not been measured. Mass estimates for classical double-mode pulsators exist, but the validity and accuracy of non-linear multimode models is still debated (see, e.g., Molnár et al. 2015). Linear models of high-degree modes offer a new (although still model-dependent) alternative. Furthermore, the [Fe/H] indices RR Lyrae stars can also be estimated based on their light curve shapes (Jurcsik & Kovács, 1995).

Globular clusters include other variable stars as well. The blue stragglers of the clusters may be SX Phe pulsators, thus giving us seismic information about these strange objects (Gilliland et al. 1998; Nemec et al. 2017). Since these stars are positioned above the turnoff point, high-quality time series photometry could be collected for them, which can then be fitted with seismic models.

Oscillations and/or pulsations in M dwarfs have been predicted at very low amplitudes, but are yet to be detected (Rodriguez-López, 2019). Discovery of oscillating M dwarfs would open up an entirely new field, where models of fully convective stars can be tested against seismic signatures. This would be of great relevance not only because many M dwarfs are planet hosts, but because evolutionary models of M dwarfs are still not reproducing the observations satisfactorily. M dwarf populations of the clusters are likely too faint for this purpose, but foreground stars could still reveal whether the predictions for their oscillations are correct.

**Searching for new variables**

Some of the clusters have few or no variable stars identified. A high-precision time-domain survey will be ideal to discover several new variables in the proposed clusters. A thorough search can reveal previously unknown targets even in clusters that have been surveyed from the ground. Many new low-amplitude variables were found in M4 from the K2 observations (Wallace et al., 2019b), as well as in other clusters observed by Kepler (Molnár et al., in prep; Kalup et al., in prep).

## 4. Comparison with Kepler, HAYDN and Earth 2.0 (ET)

To date only the Kepler and TESS missions have collected continuous, high-precision photometry for globular clusters, and between those missions only Kepler had the angular resolution to produce useful data. Here we summarize the experiences and lessons learned from the Kepler mission. The space telescope observed eight globular clusters during the K2 mission, ranging from M4, the closest cluster to the Sun, to Terzan 5, a faint and distant target. Some of the clusters were only covered partially.

Most globular clusters are compact and dense objects and separating their stars requires high angular resolution. Kepler's pixel resolution was 1/40th of Roman's planned resolution, and yet, we were able to extract precise photometry for several variable stars in M4, NGC5897 and the intrinsically dense cluster M80, whereas analysis of the remaining clusters is under way.

The photometric precision of Kepler made it possible to detect seismic signals in the giant population of M4, including the lower and upper RGB, the red HB and the AGB, down to about G ~ 14.5 mag. However, all other clusters are farther away and therefore any possible seismic detections are limited to the upper RGB and AGB, with the HB brightness level being at 16 mag or below. Here we can only rely on RR Lyrae masses.

Roman, in comparison, will easily access stars and provide us with high-precision down to 18-20 mag, and therefore will be able to provide high-precision photometry down to the base of the red giant branch for the brighter clusters. It will have a much higher resolution than Kepler, therefore contamination and blending will be much less of an issue.

Where Roman will be at a disadvantage compared to Kepler is the handling of the brightest stars in the clusters. For Kepler it was possible to extract useful data up to 1-2 mag in the V band thanks to various pipelines developed by the community. One technique is the "halo" photometry, where the light variations are reconstructed from the deep PSF and scattered-light halo around the saturated pixels. The method was developed by Pope et al. 2019. The same technique was implemented in the TESS Data for Asteroseismology photometric pipeline for the TESS space telescope (Handberg et al., 2021). It will be desirable to extend the photometric range of Roman as well.

Other proposed missions that are expected to collect time-series photometry of globular clusters in a similar manner are the European HAYDN[1] (Miglio et al. 2021) and the Chinese Earth 2.0 (ET) projects (Ge et al., 2022). HAYDN is an ESA Voyage 2050 mission proposal that would specifically target dense stellar fields, clusters as well as the Bulge, for asteroseismic purposes. The proposal envisages a telescope similar to Kepler in size, but with a much better angular resolution, close to that of Roman (0.25"/px). In contrast, Earth 2.0 will have a pixel scale similar to that of Kepler and will use a 30 cm telescope to survey the Bulge for microlensing. These three missions have the potential to complement each other: Roman and HAYDN collecting comparable, high-quality IR and optical data, with Roman also providing very deep photometry, whereas Earth 2.0 focusing on seismic data for stars at the bright end.

---

[1] http://www.asterochronometry.eu/haydn/

## 5. Recommendations

In order to enrich the stellar astrophysics portfolio of the Nancy Grace Roman Space Telescope, we recommend the following.

**5.1 Point Roman towards globular clusters in dedicated observing runs**

As we outlined above, the scientific benefits of observing one or more globular clusters as part or an extension of the Galactic Bulge Time Domain Survey would be enormous, potentially warranting a dedicated observing run (just like the dedicated Bulge or supernova campaigns in the K2 mission). This can happen either during the nominal mission or during a mission extension: the latter may allow for larger flexibility to place the field-of-view around Baade's Window. Our two highest-priority targets are NGC 6522 and NGC 6528 which can be covered in a single pointing and are well within the Window, allowing for a very dense stellar background that's needed for microlensing surveys.

**5.2 Implement faster cadence and longer exposures to detect fast variations and solar-like oscillations**

The currently proposed 15 min cadence will have a Nyquist frequency of 555 microHz, which limits the sensitivity of Roman to red giant oscillators. We recommend increasing the cadence rate to 2 min for at least a dedicated observing run, similar to the fast cadence mode of TESS. This will raise the Nyquist limit to above 4000 Hz, which would allow the detection of faster variations in SX Phe stars, or maybe even solar-like oscillations on the subgiant branch and the main sequence. The latter will also require longer exposure times than the <60 s indicated in Penny et al. (2019) to maximize the photon count.

**5.3 Develop photometric techniques to extract highly saturated stars**

The K2 mission had an extremely wide brightness range spanning 20 magnitudes thanks to the wide variety of science cases and processing methods developed for the mission. Some of these, such as the halo photometry, was aimed at extracting or reconstructing the brightness variations of very bright stars. Development and/or adoption of such methods will be very beneficial as it can extend the photometry 5-10 magnitudes above the saturation limit, and will allow us to sample the stars in the upper RGB and AGB of the proposed clusters, too.

**Acknowledgements.** This research was supported by the `SeismoLab' KKP-137523 Élvonal grant of the Hungarian Research, Development and Innovation Office (NKFIH). M.J. gratefully acknowledges funding of MATISSE: *Measuring Ages Through Isochrones, Seismology, and Stellar Evolution*, awarded through the European Commission's Widening Fellowship. This project has received funding from the European Union's Horizon 2020 research and innovation programme. This research has made use of the SIMBAD database, operated at CDS, Strasbourg, France, and NASA's Astrophysics Data System Bibliographic Services.